\documentclass{emulateapj}
\usepackage{epstopdf}
\usepackage{amsmath, amsthm, amssymb} 
\usepackage{fancybox}
\usepackage{graphicx} 
\usepackage{color} 
\usepackage{epsfig}
\newcommand{\be}{  \begin{eqnarray} }
\newcommand{\ee}{  \end{eqnarray} }

\def\spose#1{\hbox to 0pt{#1\hss}}
\def\lta{\mathrel{\spose{\lower 3pt\hbox{$\mathchar"218$}}
     \raise 2.0pt\hbox{$\mathchar"13C$}}}
\def\gta{\mathrel{\spose{\lower 3pt\hbox{$\mathchar"218$}}
     \raise 2.0pt\hbox{$\mathchar"13E$}}}
\begin{document}

\shorttitle{Thermal Tidal Inflation of Hot Jupiters }
\title{Relationship between thermal tides and radius excess}
\author{Aristotle Socrates}
\affil{Institute for Advanced Study, Einstein Drive, Princeton, NJ 08540}

\begin{abstract}
Close-in extrasolar gas giants -- the hot Jupiters -- display 
departures in radius above the zero-temperature solution, the radius excess,
that are anomalously high.  The radius excess of hot Jupiters follows a relatively close relation
with thermal tidal tidal torques and holds for $\sim 4-5$ orders of magnitude in
a characteristic thermal tidal power in such a way that is consistent with basic 
theoretical expectations.  The relation suggests that thermal tidal torques determine the global thermodynamic and spin state of the hot Jupiters.  On empirical grounds, it is shown that theories of hot Jupiter inflation that invoke a constant fraction of the stellar flux to be deposited at great depth are, essentially, falsified.   
\end{abstract}
\keywords{planets: hot Jupiters}
\section{Introduction}

The radii of hot Jupiters are anomolously
larger than their zero temperature radii, $R_0$ (Bodenheimer et al. 2003; 
Arras \& Bildsten 2007).  The powerful stellar radiation field 
serves as an ample source of energy which may inflate
hot Jupiters to their current radii (Guillot \& Showman 2002).  The
major theoretical challenge is in finding a mechanism that can 
transfer a small amount ($\lesssim 1\%$) of the stellar insolation 
to  great depth.  
Current ideas include thermal tides (Arras \& Socrates 2009a,b; 2010),
delayed gravitational contraction (Ibigui et al. 2010) and 
ohmic dissipation due to magnetic activity on the heavily irradiated 
surface (Batygin \& Stevenson 2010).

It seems that there is an relation between observed radius 
excess $\Delta$ defined as
\be
\Delta \equiv \frac{R_p-R_0}{R_p},\nonumber
\ee
where $R_p$ is the measured planet radius,
and a characteristic thermal tidal power
$L_{\rm thT}$ defined as
\be
L_{\rm thT}\equiv  \frac{\sigma_{_{\rm SB}}}{c_p}n^2\,T^3_{\rm eq}\,\nonumber
R^4_p
\ee
where $\sigma_{_{\rm SB}}$ is the Stefan-Boltzmann constant, $c_p$ is the 
specific heat, $n$ is the mean motion and $T_{\rm eq}$ is its equilibrium 
surface temperature.

 The radius excess $\Delta$ is determined by a balance of heating and cooling. 
 In the thermal tide scenario, the thermal tide torque is balanced by the gravitational torque. While the thermal tide torque is non-dissipative, the gravitational tide torque necessarily is the result of 
 dissipation. Torque balance in steady state then implies continuous heating as a result 
 of gravitational tidal dissipation.  It follows that
 the planet radius is determined by a balance of this steady state heating with cooling at the 
 photosphere. 
 The ultimate source of energy is the stellar radiation field, which performs work by 
 moving material across the tidal potential, near the surface of the irradiated planet (Arras \& Socrates 2009a; 2010).

The plan of this paper is as follows:  In \S\ref{s: theory} we motivate the 
need for thermal tidal torques in hot Jupiters, compute their characteristic
values for both equilibrium and dynamical thermal tides as well as the 
the characteristic thermal tidal power, all in terms of physical quantities 
that can be observationally inferred.  The thermal tide - radius excess relation 
is presented in \S\ref{s: relation}.   A brief discussion is contained in 
\S\ref{s: discussion}.  A brief summary is given in \S\ref{s: conclude}.

\section{Theory of Thermal tides in hot Jupiters}\label{s: theory}

The radiation field of a star forces the atmospheres of planets in 
orbit.  The response of a planetary atmosphere to this forcing is 
referred to as the ``thermal tide."  Planets that spin
asynchronously experience a sharp change in thermal forcing 
during twilight hours and consequently, a broad spectrum of 
tidal harmonics are excited.    For circular orbits, the semi-diurnal $(\ell=m=2)$
harmonic of the forcing leads to a thermal tidal bulge that can couple to 
the leading order term of the gravitational tidal potential. The atmosphere's
finite thermal inertia leads to a lag in the response with respect to the 
forcing, which allows the tidal potential to exert a net torque on the thermal 
tide during the heating cycle.  Since the thermal tidal bulge peaks in the morning and
after sunset, but before midnight, the acceleration increases the planet's rotational 
energy.   

The picture described above has a long history.  On the basis of the measured 
magnitude and phase of Earth's thermal tide, Thompson (Lord Kelvin) computed Earth's 
thermal tidal torque and found it to be approximately equal to one tenth the dissipative
lunar tidal torque (Thompson 1882;  see also Munk \& Macdonald 1960).  Due to its 
heavy atmosphere and correspondingly large thermal tides,
Gold \& Soter (1969) hypothesized that the current spin state
of Venus results from a  balance of thermal and gravitational tidal torques so that
its current state of spin can be maintained over the age of the solar system.  

In what follows, we motivate the importance of thermal tides and thermal tidal 
torques in hot Jupiters, within the context of understanding their
inflated radii.

\begin{figure}
%\begin{center}
\epsscale{1.2}
\plotone{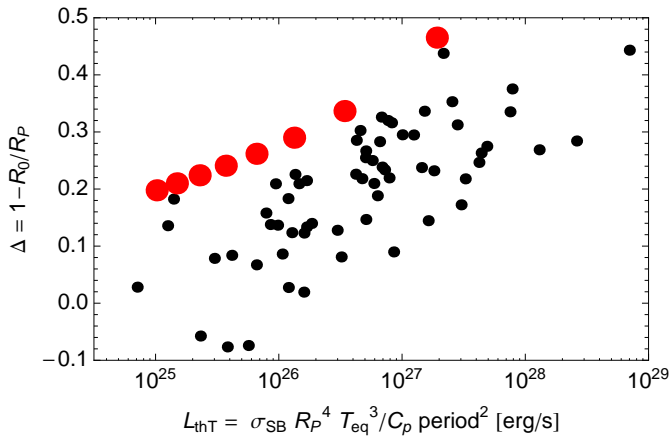}
\caption{  The $\Delta-L_{\rm thT}$ relation for a sample of 63 hot Jupiters
with $M\sin i > 0.5\,M_J$.
An approximation for the zero temperature radius $R_0$ is obtained by using the 
cooling models  of Baraffe et al. (2010) for core-less planets with $Z=0.02$
in the absence of irradiation at an age of $7$ Gyrs.   Large
red points are theory, computed using the Gold-Soter approximation for 
a Jupiter-mass, core-less $Z=0.02$ planet whose optical absorption opacity
$\kappa_{\star}=0.003$ (Arras \& Socrates 2009a).  The tidal quality factor of the planet is chosen such 
that $Q=10^5$ for the Jupiter-Io interaction 
and the constant lag time model of Hut (1981) is employed
in order to compute the gravitational tidal torque and equilibrium dissipation rate.  }
\label{fig: correlation}
%\end{center}
\end{figure}

\subsection{tidal power as a source of dissipation}

The characteristic value for the maximum instantaneous tidal power for
a uniform spherical body of radius $R$ in a Keplerian 
orbit about a central source of mass $M$ 
at a heliocentric distance $D$ is given by
\be
L_{\rm max,T}\approx\frac{c_{_{\rm T}}^5}{G}
\label{e: tide_max}
\ee
where the characteristic tidal speed $c_{_{\rm T}}$ is given
by
\be
c_{_{\rm T}}\equiv R\,\sqrt{\frac{G\,M}{D^3}}
\ee
and $G$ is Newton's constant\footnote{Note the similarity 
of eq. \ref{e: tide_max} with maximum luminosity of any 
object in the Universe
\be
L_{\rm max}=\frac{c^5}{G}. \nonumber 
\ee
}.
The expression above is obtained by taking 
the peak energy of the equilibrium tidal response to dissipate
with a timescale given by the inverse of 
$\sqrt{G\,M/D^3}$.  

Most hot Jupiters are in nearly circular orbits. In that case, 
$n=\sqrt{GM/a^3}$ is the mean motion and only the synchronization 
tide can lead to tidal dissipation and therefore, the forcing 
frequency is given by the difference of the spin frequency and
mean motion.  If the planet is rapidly spinning such that the characteristic 
tidal forcing frequency in eq. \ref{e: tide_max} is taken to be the 
mean motion $n$, then eq. \ref{e: tide_max} may be written as
\be
L_{\rm max,T}\approx 10^{34}\,P^{-5}_{3}\,R^5_{10}
\ee
where $P_3$ is the orbital period $P_{\rm orb}$ in units 
of 3 days and $R_{10}$ is the planet's radius, $R_p$, normalized
to $10^{10}$ cm.  The power required to maintain the radii of 
heavily inflated hot Jupiters is of order $\approx 10^{28}$ erg/s
and therefore, only a small fraction of the total available tidal power
is required to produce inflated hot Jupiters in steady-state.

\subsection{the role of thermal tides}

Stellar irradiation creates a tidal bulge
that leads maximum forcing at noon, where the phase shift
results from thermal inertia.  The tidal gravitational field of the 
star torques the thermal tide, accelerating the planet away
from a state of synchronous spin.  As the rate of spin increases, the 
amount of energy absorbed per cycle decreases and consequently, so does
the magnitude of the thermal tidal quadrupole.  The thermal tide torque 
eventually comes into 
balance with the usual dissipative gravitational tidal spin-down torque. 
In steady-state, the planet's spin is asynchronous and therefore, 
gravitational tidal power is continuously dissipated.  
The ultimate source of energy is the starlight of the primary, which performs work by 
moving material across the tidal potential (Arras \& Socrates
2009a; 2010). 

A fundamental underlying assumption of this picture is that dissipation 
of the gravitational tide takes place at great depth, where the pressure scale height is 
comparable to the radius of the planet.  

Also note that, in the absence of thermal tidal toques, gravitational tidal torques are likely
to synchronize the planet spin extremely fast if the relative strength of tidal dissipation 
in hot Jupiters is comparable to what is commonly inferred from the Jupiter-Io interaction
(Goldreich \& Soter 1966; Socrates et al. 2012).

\subsection{spin equilibrum and steady tidal power:  Gold-Soter approximation}

Balance between the gravitational tidal torque and 
the thermal tidal torque determines the planet's spin 
rate and therefore, determines the tidal forcing frequency $\omega$.  
For a circular or nearly circular orbit, torque balance 
is obtained by equating the respective quadrupoles 
induced by thermal and gravitational forcing.
The portion of the gravitationally excited quadrupole responsible
for dissipation and secular evolution is given by

\be
q_{{\rm grav}} = \frac{n^2R^5_p}{Q_{_{\rm J}}\omega_{_{\rm J}}\,G}\omega
\label{e: q_grav}
\ee
where the above expression for the tidal quality factor $Q$ reflects
the frequency dependence of the constant lag time model of Hut (1981).
Here $Q_J$ and $\omega_J$ is the tidal quality factor($\approx 10^5$; Goldreich
\& Soter 1966) and forcing frequency ($\approx 10$ hours) of the Jupiter-Io 
interaction.  Note that we are, equivalently, setting the lag time $\tau\approx 0.1$
s, consistent with the resonant configuration of the Galilean satellites (Socrates
et al. 2012; Leconte et al. 2010; see also Socrates \& Katz 2012).

The Gold-Soter quadrupole due to thermal forcing is aproximately given by
\be
q^{_{\rm GS}}_{\rm th}=\frac{\Delta M\,R^2_p}{t_{\rm th}\omega}
=\frac{\sigma_{_{\rm SB}}}{c_p}\frac{R^4_pT^3_{\rm eq}}{\omega}
\label{e: q_gs}
\ee
where $\Delta M$ and $t_{\rm th}$ are the mass and thermal relaxation time, respectively, 
of the absorbing layer and $t_{\rm th}$.  In terms of physical quantities, the 
the thermal time may be written as 
\be
t_{\rm th}\equiv \frac{c_p}{\sigma_{_{\rm SB}}}\frac{\Delta M}{R^2_pT^3_{eq}}.
\ee

By equating eq. \ref{e: q_grav}  with eq. \ref{e: q_gs}, the 
equilibrium forcing frequency becomes
\be
\omega_{_{\rm GS}} & = & \sqrt{\frac{\sigma_{_{\rm SB}}\,Q_{\rm J}\,\omega_{\rm J} }{c_p}
\frac{T^3_{\rm eq} }{n^2\,R_p}}\nonumber\\
& \approx & \frac{2\pi}{2 \,{\rm days}}\,\,T^{3/2}_{2000}\, P_{4}\,R^{-1/2}_{10}\, Q^{1/2}_5\,
\ee
where $T_{2000}$, $P_4$, $R_{10}$, $Q_5$ are the equilibrium 
temperature in units $2000$ K, the orbital period normalized to 
four days, dimensionless planet radius normalized to $10^{10}$ cm and 
the tidal quality factor of the Jupiter-Io interaction, normalized to $10^5$.

In the equilibrium -- Gold-Soter -- approximation, the tidal dissipation rate resulting from the balance of
thermal and gravitational tidal torques, which we refer to as the 
thermal tidal power $L^{_{\rm GS}}_{\rm thT}$ is given by
\be
L^{_{\rm GS}}_{\rm thT} & = &\omega_{_{\rm GS}}\,n^2\,q_{\rm grav}=\omega^2_{_{\rm GS}}\frac{n^4\,R^5_p}{G\,Q_{\rm J}\,\omega_{\rm J}}\nonumber\\
& \approx & \frac{\sigma_{_{\rm SB}}}{c_p}\,n^2\,T^{3}_{\rm eq}\,R^4_p\nonumber\\
& = & 1.5\times 10^{28} \,P^{-2}_4\, T^3_{2000}\,R^4_{10}
\,{\rm erg/s}.
\label{e: GS_lum}
\ee

\subsection{spin equilibrium and steady tidal power: dynamical thermal 
tides}

Consider the computed thermal tide quadrupole from Arras and Socrates (2010;
see their figure 5)
for the case with outgoing radiation boundary condition.  We imagine
an evolutionary scenario in which the planet was born spinning 
rapidly and was then placed in its current orbit via some migration 
mechanism.  As it spins
down, the forcing frequency approaches the g-mode ``bump'' in 
the thermal tidal quadrupole, which is responsible for arresting any further
spin evolution.  The first spin up feature of the 
g-mode bump is located near the cutoff frequency $\omega_0$
\be
\omega_0=\frac{H}{R_p}N
\ee  
where $H$ is scale height of the isothermal layer that lies above the 
convection zone and $N$ is the Brunt-Vaisalla frequency, the characteristic
rate at which the buoyancy force restores motion.  

It is worth noting that the
cutoff frequency $\omega_0$ is close to the mean motion $n$ of hot Jupiters.  If
$\omega_0\ll n$, the quadrupole moments associated with low radial order
gravity waves would be small and therefore, so would the quadrupole of 
the thermally-forced response.  From this perspective, it is somewhat of a co-incidence that
dynamical thermal tides in the hot Jupiters can lead to significant
spin-up tidal torques.

%\begin{figure}[t]
%\epsscale{1.2}
%\plotone{framework.png}
%\caption{Framework for thermal tidal theory of hot Jupiter radius excess (Arras
%\& Socrates 2009a).  At a given 
%spin and tidal forcing frequency, the fluid response resulting from thermal 
%forcing is computed via a linear fluid mechanics calculation.  The thermal 
%tidal quadrupole is then obtained by integrating the fluid response over the
%entire planet.  Spin equilibrium is then determined by equating the computed 
%thermal tidal quadrupole to the gravitational tidal quadrupole.  Knowledge
%of the spin frequency allows to obtain the steady state dissipation rate due 
%to gravitational tidal friction, since the planet spins asynchronously.  Along
%with the outer insolation boundary condition, the equilibrium radius of the 
%planet is determined by a planetary structure code.  There are three input parameters or
%assumptions in the theory: (i) the value of the optical absorption opacity (ii) dissipation of the gravitational tide occurs at great depth (iii) the associated tidal lag time
%is comparable to that inferred from the Jupiter-Io interaction.}
%\label{fig: framework}
%\end{figure}

The spin-up quadrupole due to the first feature in 
the {\it g}-mode bump may be expressed as 
\be
q^g_{\rm th}=q^{_{\rm GS}}_{\rm th}(\omega=\omega_0)f(\omega,\omega_0)
\ee
where $f$ can be approximated by a sharply-peaked 
Gaussian of the form
\be
f(\omega,\omega_0)\simeq \beta\, e^{-(\omega-\omega_0)^2/\Gamma^2}
\ee
where $\beta\approx 0.1$ quantifies the amount by which gravity waves
near the cut-off frequency do not perfectly overlap with both 
thermal and gravitational forcing.  The width of the Gaussian is 
narrow so that $\Gamma\ll\omega_0$.

Spin equilibrium is determined by setting $q^{g}_{\rm th}=q_{\rm grav}$, which 
gives
\be
e^{-(\omega-\omega_0)^2/\Gamma^2}\simeq\left[\frac{c_p}{\sigma_{_{\rm SB}} G}
\frac{n^2R_p\omega^2_0}{\beta\, T^3_{\rm eq}
Q_{_{\rm J}}\,\omega_{_{\rm J}}   }  \right]\,\frac{\omega}{\omega_0}.
\label{e: torque_dyn}
\ee
The right hand side must be less than unity in order 
for torque balance to be satisfied.  For the physical parameters that describe the 
hot Jupiter population, torque balance is only possible for $Q_{_{\rm J}}\gtrsim
10^5$.  If there is some level of sub-surface reflection the excited gravity waves
are quantized and their response is therefore, characterized by a Lorentzian profile
near resonance, implying that the induced quadrupolar response can be much larger.  Consequently,
values of $Q_{_{\rm J}}\approx 10^4$ or $\tau\simeq 1$ s
are then possible for a steady-state spin equilibrium (Arras \& Socrates 2010).

Due to the sharpness of the Gaussian, the forcing frequency in
spin equilibrium, given by eq. \ref{e: torque_dyn}, is 
pinned to a value that is close to the cutoff frequency
\be
\omega\approx \omega_0.
\ee
By following the steps in eq. \ref{e: GS_lum}, a similar 
expression for thermal tidal power, $L^{_{\rm dyn}}_{\rm thT}$ is derived due to the presence
of dynamical tides
\be
L^{_{\rm dyn}}_{\rm thT}\approx  1.2\times10^{28}\,\omega^2_{2}\,P^{-4}_{4}\,
R^5_{10}\,Q^{-1}_5\,{\rm erg/s}
\label{e: L_th_dyn}
\ee
where $\omega_2$ and $Q_5$ are the semi-diurnal
 forcing frequency normalized to $\left(2\pi/2 \,{\rm days}
\right)$ and the tidal quality factor of the Jupiter-Io interaction, normalized to 
$10^5$.

The thermal tidal dissipation rate, eq. \ref{e: L_th_dyn} resulting from the dynamical excitation of 
gravity waves, is similar in form and magnitude to that resulting from 
the Gold-Soter approximation, given by eq. \ref{e: GS_lum}.  From here on, when comparing 
with data, we employ the Gold-Soter approximation since (i) it is simple (ii) in terms of
spin equilibrium and persistent tidal luminosity,  it roughly agrees with 
the more realistic dynamical tide results of Arras \& Socrates (2010) (iii) the theory of dynamical 
thermal tides is currently in a state of development.

\section{Thermal tide-radius excess relation}\label{s: relation}

The fractional radius anomaly $\Delta$, given by
\be
\Delta \equiv 1-R_0/R_p,
\ee
where $R_0$ is a fiducial radius that is close
the zero temperature solution depends upon planet mass and composition.
The value for $R_0$ is obtained by using the on-line table provided
by Baraffe et al. (2008) by choosing 
choosing $Z=0.02$ with no irradiation and their 
largest computed age of $7$ Gyr. 

The primary reason for utilizing the models of Baraffe et al.
for the values of $R_0$, rather than the actual zero temperature
or isothermal radius is that they are publicly available.  Their solutions
serve as a fiducial value of low planetary entropy that depends on
planet mass.  It follows that any observed departure in planet radius
$R_p$ above $R_0$ necessitates a source of thermal energy from within 
the planet.

The sample of hot Jupiters is taken from the Exoplanet
Data Explorer, located on the world wide web.
Only hot Jupiters with measured mass, radius, effective temperature and 
orbit that have measurements  i.e., a sizable fraction of the
total sample, are considered.  Furthermore, the sample is limited to hot Jupiters with masses 
$M>0.5M_J$.  All of
the planets considered have both transit and radial velocity
measurements, for a total of 63 objects.

In figure \ref{fig: correlation} we plot radius anomaly
$\Delta$ against the dissipation rate given by eq. (\ref{e: GS_lum}).
There is a clear relationship.  Any level of deep internal energy 
generation 
increases the central entropy, lifts the degeneracy and 
expands the planetary radius.  Figure \ref{fig: correlation} 
indicates that the radius anomaly, or 
departure from the zero temperature radius, increases with 
an increasing internal dissipation rate that is correlated to the
characteristic thermal tidal dissipation rate $L_{\rm thT}$.  That
is, figure \ref{fig: correlation} lends strength to the hypothesis that
thermal tidal torques power the core luminosities 
of hot Jupiters.

Figure \ref{fig: correlation} implies that radius excess $\Delta$, rather than 
planet radius $R_p$ is more sensitive in ascertaining the actual internal luminosity 
of a given planet and is therefore, a more useful parameter.  The reason for this is 
straightforward:  the mass-radius relation of cold spheres (e.g., Zapolsky \& Salpeter 1969) 
indicates that -- for a fixed composition -- radius is approximately independent of mass.  
Thus, the gravitational binding energy
and approximately, the Fermi energy, vary by nearly two orders of magnitude.  
As a result, the energy and power requirements to lift the degeneracy and inflate the 
radius are far more severe for massive objects.  The relatively small scatter in 
figure \ref{fig: correlation} in comparison to similar figures that utilize planet radius 
(e.g., figure 1 of Demory \& Seager 2011; figure 1 of Lauglin et al. 2011; figure 1
of Fortney et al. 2011) is partially due to this fact.

\section{discussion}\label{s: discussion}

Figure \ref{fig: correlation} is important irrespective of whether or not 
it supports the thermal tide model of Arras \& Socrates.  It indicates that
the internal luminosity of the hot Jupiters varies by $\sim 4-5$ orders of 
magnitude.  Such a conclusion can be indirectly inferred from e.g., recent
work by Spiegel \& Burrows (2013), independent of the actual source of 
thermal power at great depth.

With the exception of the thermal tide scenario advocated here, nearly every plausible model 
of steady-state hot Jupiter inflation invokes or presumes some fixed fraction of absorbed stellar energy 
to be redistributed to great depth (e.g., Guillot \& Showman 2002;  Batygin \& Stevenson 
2010; Youdin \& Mitchell 2012; Laughlin et al. 2012).  The stellar flux varies like
$1/D^2$ and therefore, so does -- for fixed planet radius and stellar type -- the available amount 
of energy absorbed.  Largely due to the fact that hot Jupiters -- by definition -- possess orbital separations $0.01 {\rm AU}\leq D \leq 0.1{\rm AU}$,
the spread in the available energy in starlight varies by approximately 
two orders of magnitude.  Since the observed spread in radii requires a corresponding
spread in steady state power of $\sim 4-5$ orders of magnitude, theories that utilize 
a constant fraction of the absorbed stellar light are then, effectively, falsified.

In general, the amplitude of a tidal response is $\propto 1/D^3$ and therefore, the 
energy in the tidal interaction, which is proportional to the square of the tidal amplitude, 
is  $\propto 1/D^6$.  For the case of thermal tides, the tidal response is forced by the 
stellar radiation flux and is therefore $\propto 1/D^2$.  The tidal interaction, which 
leads to orbital and spin evolution, is given by the product of the thermal tide quadrupole 
and the tidal potential and is thus $\propto 1/D^5$.  Consequently, the expected spread
in thermal tidal power for the hot Jupiters is roughly $5$ orders of magnitude, consistent
with what is required to inflate their radii.

The hot Jupiters sit deep within a steep tidal potential ($\propto 1/D^3$) that
leads to even more rapidly varying tidal interaction potentials ($\propto 1/D^6$ for gravitational 
tides and $\propto 1/D^5$ thermal tides).  Relatively modest displacements in 
orbital separation leads to relatively large changes in the internal tidal luminosity.  In 
a way, the radius excess phenomena displayed by the hot Jupiters can be thought of
as a manifestation of, otherwise degenerate, gaseous self-gravitating bodies probing
deep and rapidly varying interaction potentials.\footnote{Black widow pulsar systems (Fruchter et al. 1988) for example, have physical properties that are quite similar to those of hot Jupiter systems.  The sub-stellar companions to black widow pulsars are inflated as well and are thought to be powered by some type of tidal mechanism (Applegate \& Shaham 1994).  Rather than optical photons as in the case of hot Jupiter hosts, the radiative output of the central source is a relativistic outflow composed of some mixture of particles, photons and magnetic field, which is 
ultimately derived from the spin-down power of the neutron star. }

\section{Summary}\label{s: conclude}

An observed relationship between radius excess $\Delta$ and a characteristic thermal tidal power
$L_{_{\rm thT}}$ for the hot Jupiters is given in figure \ref{fig: correlation}.  It
indicates that, in order to produce the observed trend in radius excess $\Delta$, the internal 
luminosity must vary by $\sim 4-5$ orders of magnitude for the population of inflated hot 
Jupiters.  This empirical requirement lends support to {\it any} tidal scenario that inflates the hot 
Jupiters steadily over the age of the Galaxy and therefore, lends support to the steady-state thermal 
tide scenario of Arras \& Socrates (2009a; 2010).  

Furthermore, the empirical requirement that a large spread of inferred internal luminosity is required
to explain the trend in radius excess $\Delta$ from figure \ref{fig: correlation}
falsifies theories of hot Jupiter inflation that invoke, presume or calculate that a fixed fraction 
of the absorbed stellar power is transferred to great depths. 

Both equilibrium and dynamical thermal tides approximately yield comparable values for 
equilibrium thermal tidal power and planetary spin rate.  The role of thermal tides can be 
viewed as allowing for the steady dissipation of tidal energy that is continuously replenished
by the stellar radiation field as it performs work by moving matter across the tidal potential (Arras \& Socrates 2009a; 2010).

\acknowledgements I thank P. Arras for extensive conversations and for assistance.  
Subo Dong, Andy Gould and Scott Tremaine provided helpful suggestions.  AS 
acknowledges support from a John N. Bahcall Fellowship at the Institute for 
Advanced Study, Princeton.

\end{document}